\documentclass{INTERSPEECH2023}
\usepackage{subcaption}

\interspeechcameraready

\title{Supervised Contrastive Learning with Nearest Neighbor Search for Speech Emotion Recognition}
\name{Xuechen Wang$^1$, Shiwan Zhao$^*$, Yong Qin$^{1, \dag}$}
\address{
  $^1$Nankai University, Tianjin, China
  }
\email{shirleywxc0103@mail.nankai.edu.cn, zhaosw@gmail.com, qinyong@nankai.edu.cn}

\begin{document}
\maketitle

\renewcommand{\thefootnote}{}
\footnotetext{* Independent researcher.}
\footnotetext{\dag Corresponding author.}

\begin{abstract}
Speech Emotion Recognition (SER) is a challenging task due to limited data and blurred boundaries of certain emotions. In this paper, we present a comprehensive approach to improve the SER performance throughout the model lifecycle, including pre-training, fine-tuning, and inference stages. To address the data scarcity issue, we utilize a pre-trained model, wav2vec2.0. During fine-tuning, we propose a novel loss function that combines cross-entropy loss with supervised contrastive learning loss to improve the model's discriminative ability. This approach increases the inter-class distances and decreases the intra-class distances, mitigating the issue of blurred boundaries. Finally, to leverage the improved distances, we propose an interpolation method at the inference stage that combines the model prediction with the output from a k-nearest neighbors model. Our experiments on IEMOCAP demonstrate that our proposed methods outperform current state-of-the-art results. 

\end{abstract}

\section{Introduction}
The recognition of emotions has emerged as a significant aspect in the field of human-computer interaction. Speech, being a rich source of emotional cues conveyed through various attributes such as pitch, frequency, speed, and accent, has received considerable attention in this regard. With the development of artificial intelligence technologies, Speech Emotion Recognition (SER) has been applied in a broad range of domains such as online education, human customer service, psychological health, and entertainment. Nevertheless, the recognition of emotional categories from speech utterances presents a daunting challenge due to their abstract and complex nature, coupled with limited data availability.

Recent advances in deep learning have made it the primary method for SER. Specifically, Badshah et al. \cite{7883728} proposed a framework based on Convolutional Neural Networks (CNNs) to predict emotions. Satt et al. \cite{satt17_interspeech} used spectrograms extracted from speech as the model input directly, obtaining better performance and limiting the latency. Mirsamadi et al. \cite{7952552} used Recurrent Neural Networks (RNNs) to learn frame-level features. Kumawat et al. \cite{kumawat21_interspeech} applied Time Delay Neural Network (TDNN) to capture the temporal information and provide an utterance level prediction. Liu et al. \cite{10045019} proposed a method with attention mechanism to capture both temporal and frequency domain context information from input spectrograms. 
Despite these developments, challenges remain in this field that need to be addressed.
 
Compared to other related fields, the datasets in SER are often limited in size, making it challenging to train large-scale and robust models exclusively using SER datasets. Moreover, there is no generic pre-trained model available that can be directly applied to SER. Meanwhile, self-supervised pre-trained models such as Wav2vec2.0 \cite{NEURIPS2020_92d1e1eb} and Hubert \cite{hsu2021hubert} have achieved good performances in Automatic Speech Recognition (ASR). They have been trained with a large amount of speech data and could construct better feature embeddings for utterances. However, their application in other related fields, such as SER, is still in its initial stages. In this paper, we adopt a transfer learning method \cite{latif2018transfer} to address these challenges. Specifically, we leverage the self-supervised model wav2vec2.0 as a pre-trained model to obtain more accurate speech representations. We fine-tune the pre-trained model on a specific SER task to make the representations more suitable for the downstream task.

\begin{figure}[t]
  \centering
  \includegraphics[width=\linewidth]{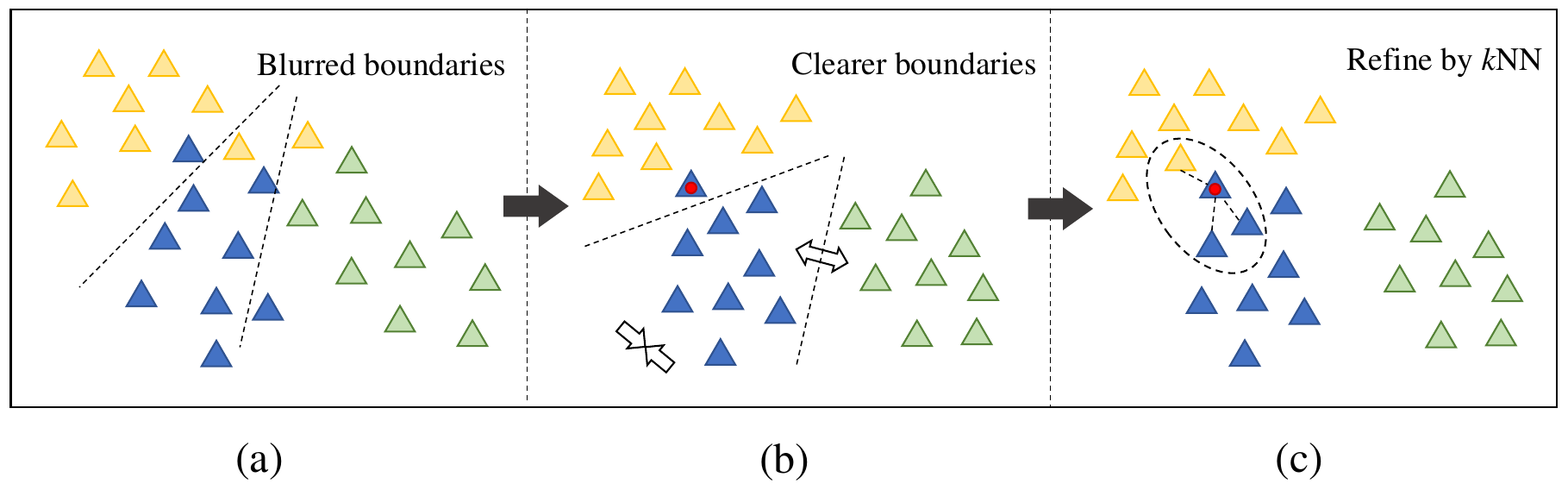}
  \caption{The overall approach throughout the entire lifecycle of SER. (a): Feature embeddings after pre-training. (b): Feature embeddings after fine-tuning with the help of supervised contrastive learning. (c): Refine predictions by interpolating model outputs with a $k$NN model in the inference stage.}
  \label{fig:KNN-Contrastive}
  \vspace{-6pt}
\end{figure}

\begin{figure*}[t]
  \centering
  \includegraphics[width=\linewidth]{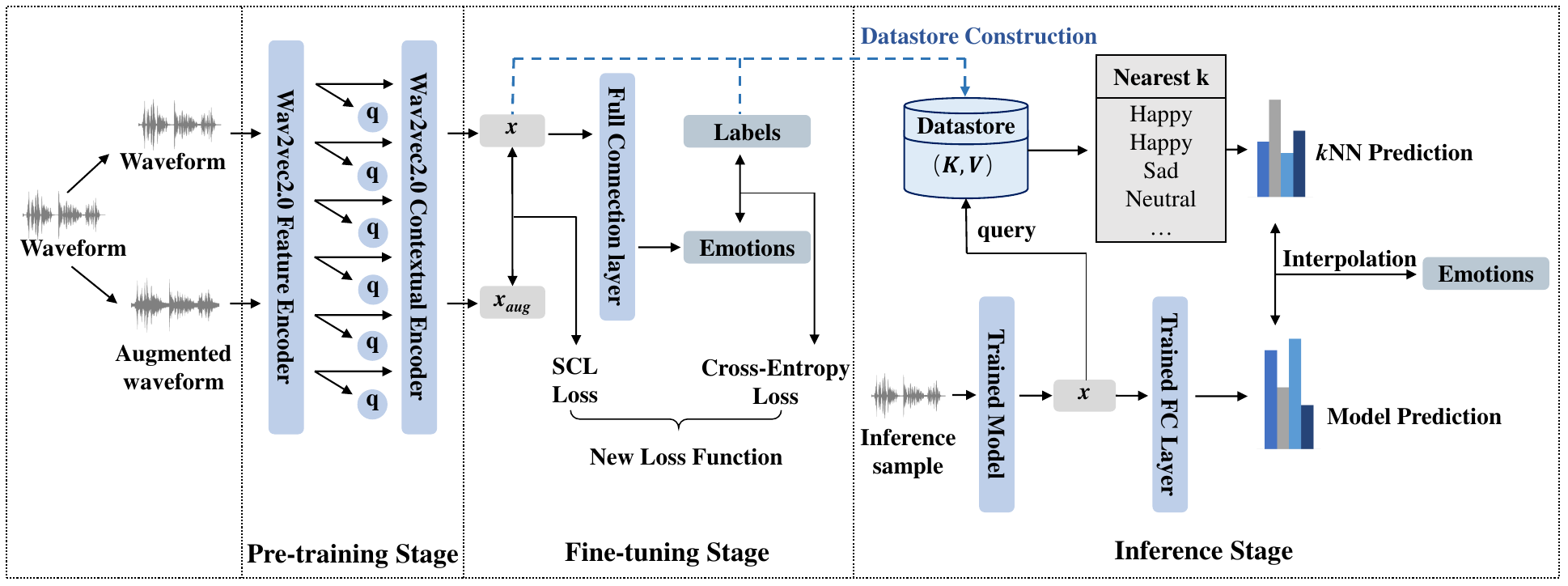}
  \caption{The overall framework of our proposed method to improve the performance of SER  throughout all stages, including the stages of pre-training, fine-tuning, and inference.}
  \label{fig:overview}
\end{figure*}

In the fine-tuning stage, previous works \cite{8282123, zhou2020transfer} used cross-entropy loss to guide the model to solve multi-class classification problems. However, some researches \cite{NEURIPS2018_f2925f97, sukhbaatar2015training} found that cross-entropy loss may result in poor generalization performance and lack robustness to noisy labels, especially when the training data is limited. In addition, due to the blurred boundaries of certain emotions, general models may struggle to distinguish these emotions, for example, resulting in the misclassification of excited emotions as neutral emotions \cite{ma18b_interspeech}. To solve the above problems, Li et al. \cite{9413910} and Lian et al. \cite{lian2018speech} introduced contrastive learning to pull samples from the same classes closer and push samples from different classes farther apart. However, such naive contrastive learning approaches could not make full use of data labels for limited data. Alaparth et al. \cite{alaparthi2022scser} applied Supervised Contrastive Learning (SCL) \cite{gunel2021supervised} loss to fine-tune the original model, which required two training stages to obtain the final predictions. In this paper, we propose a new loss function that combines cross-entropy loss and SCL loss in a weighted manner. The new objective function allows the model to learn better feature representations with increased inter-class distances and decreased intra-class distances more efficiently. This helps to address the problem of blurred boundaries and, as a result, improve the classification performance of the model.

Following the training with supervised contrastive learning, the intra-class feature representations become more compact while the inter-class feature representations become more separated. However, previous works did not fully utilize these enhanced feature representations that contain better distance information. In this study, we propose to apply the k-nearest neighbors ($k$NN) algorithm in the inference stage to fully leverage the feature distances. As illustrated in figure \ref{fig:KNN-Contrastive}, $k$NN can refine the model's predictions by interpolating between the model's output and the output of $k$NN. In the embedding space generated by the fine-tuned model, $k$NN results can be obtained by retrieving the nearest training samples. Retrieving augmented methods have been demonstrated to be effective in various Natural Language Processing (NLP) works \cite{Khandelwal2020Generalization, pmlr-v119-guu20a}. By explicitly utilizing this information during inference, we can improve the model's classification ability without additional training. To the best of our knowledge, our work is the first to leverage the retrieval mechanism to enhance the SER performance. 

In summary, our main contributions are as follows: 
\begin{itemize}
\item We propose a comprehensive framework to improve SER performance throughout the model lifecycle, including pre-training, fine-tuning, and inference stages. 

\item We combine cross-entropy loss and SCL loss in the fine-tuning stage, improving the performance with better feature representations.

\item We employ a $k$NN model to further enhance the model performance in the inference stage by leveraging the improved sample distance from SCL.

\item Our experiments on the IEMOCAP dataset demonstrate that our proposed methods outperform current state-of-the-art results, achieving 74.13\% and 75.13\% on WA and UA, respectively.

\end{itemize}

\section{Proposed Methods}
In this section, we discuss the overall architecture of our proposed method. As shown in figure \ref{fig:overview}, our approach covers the entire lifecycle of a SER pipeline, including pre-training, fine-tuning, and inference stages. We first review the pre-trained model wav2vec2.0. Then, we introduce our new learning objective combining cross-entropy loss and SCL loss in the fine-tuning stage. Finally, we present our proposed method for interpolating model outputs using the $k$NN algorithm during the inference stage.

\subsection{Pre-training Stage}

Wav2vec2.0 is a transformer-based model. It contains three modules, feature encoder module, contextual encoder module, and quantization module. The feature encoder module contains several convolution layers. The contextual encoder module is based on transformer architecture. The quantization module is used to discretize the output of the feature encoder to a finite set of speech representations via product quantization. During training, wav2vec2.0 relies on the method of self-supervised learning, using a large amount of unlabeled speech data. 

In this paper, to overcome the data scarcity issue in SER and obtain more accurate representations of utterances, we utilize wav2vec2.0 as the pre-trained model to extract features in the pre-training stage. 

\subsection{Fine-tuning Stage}
During fine-tuning, we propose a new loss function which combines cross-entropy loss and SCL loss in a weighted manner. In SCL, multiple samples belonging to the same class can be treated as positive samples to each other. To compute the SCL loss, we first define the concepts of \textit{positives} and \textit{negatives} for supervised contrastive learning in our task. 



For a set of \textit{N} instances \textit{\{$x_k$, $y_k$\}, k = 1, …, N}, $x_k$ denotes the feature embedding of a waveform and $y_k$ denotes its label, represented by one-hot code. 
A training batch consists of \textit{2N} instances, \textit{\{$x_l$, $y_l$\}, l = 1, …, 2N}, where $x_{2t}$  \textit{(t = 1, …, N)} denotes the original waveform $x_{k}$ and $x_{2t-1}$ denotes the augmented version of $x_k$ \textit{(k = 1, …, N)}. The label of an augmented waveform is the same as the original waveform, which could be expressed as $y_{2t} = y_{2t-1} = y_{k}$. Instances that have the same label \textit{y} are called \textit{positives}, and instances that have different labels are called \textit{negatives}. The loss function of supervised contrastive learning is as follows \cite{NEURIPS2020_d89a66c7}:

\begin{align}
\mathcal{L}_{scl}=\sum_{i\in I}\frac{-1}{|P(i)|}\sum_{p\in P(i)}\log\frac{\exp\left(\left(\boldsymbol{x}_i\cdot\boldsymbol{x}_p\right)/\tau\right)}{\sum\limits_{a\in A(i)}\exp\left(\left(\boldsymbol{x}_i\cdot\boldsymbol{x}_a\right)/\tau\right)},
\end{align}
where \textit{i $\in$ I = \{1, …, 2N\}} denotes the index of an instance, \textit{A(i)} denotes all indices except \textit{i} , and $\boldsymbol{x}$ denotes the feature embeddings of waveforms. \textit{P(i)} denotes all indices of the positive instances of sample \textit{i}. $\tau$ is a hyperparameter. 

After calculating the SCL loss $\mathcal{L}_{scl}$ and cross-entropy loss $\mathcal{L}_{ce}$, we could calculate the new learning objective as follows:
\begin{align}
\mathcal{L}=(1-\lambda)\mathcal{L}_{ce}+\lambda\mathcal{L}_{scl},
\end{align}
where $\lambda$ denotes a scalar weighting hyperparameter. 

\subsection{Inference Stage}
To implement the $k$NN algorithm in the inference stage, we first create the datastore with all samples in the training and validating datasets. The storage format of the datastore is as follows:
\begin{align}
(\mathcal{K},\mathcal{V})=\{(x_i,y_i), i\in\mathcal{D}\},
\end{align}
where $\mathcal{D}$ is a set of all indices of the samples from training and validating data, \textit{$x$} denotes the fixed-length representation of an input waveform computed by the trained model, and \textit{$y$} denotes its label. 

Afterward, when given an inference sample, we can retrieve its $k$ nearest neighbors from the datastore and make predictions for its label accordingly. These neighbors are calculated based on the distance in the embedding space produced by the fine-tuned model, and we adopt the \textit{Euclidean Distance} metric for this purpose.


Based on the predictions from both the $k$NN model and the trained model, we could obtain a final probability distribution of a given inference sample by a linear interpolation:
\begin{equation}
p(y|x)=\alpha p_{knn}(y|x)+(1-\alpha)p_{model}(y|x),
\end{equation}
where $\alpha$ is a scalar weighting hyperparameter, $p_{model}(y|x)$ and $p_{knn}(y|x)$ denotes the probability distributions output from the trained model and the $k$NN model, respectively. 

\section{Experiments}

\subsection{Dataset}
We evaluate the effectiveness of our proposed method on the IEMOCAP \cite{busso2008iemocap} dataset, which is widely used in SER. The dataset includes various emotions such as happy, angry, neutral, sad, surprised, excited, fearful, frustrated, disgusted, and so on. To be consistent with prior works, we conduct experiments on a subset of four emotions, namely, angry, happy, sad, and neutral, where the original happy category and excited category are merged as the happy category. In total, there are 5,531 utterances, comprising 1,103 angry, 1,636 happy, 1,084 sad, and 1,708 neutral utterances. 

\subsection{Experimental Settings}
In this paper, we use the publicly available pre-trained model, \textit{Wav2Vec 2.0 Base}\footnote{https://github.com/facebookresearch/fairseq/tree/main/examples/
wav2vec}, developed by Facebook AI. It is trained on the LibriSpeech dataset \cite{7178964} through self-supervised learning without any fine-tuning. We apply 10-fold cross-validation to evaluate our results, leaving two speakers out for each fold, one for validating and the other for testing. To retain as much information as possible and save computing resources, we set the max-length of all utterances to 7.5 seconds. The batch size is set to 12 and the max training epoch is set to 150. We choose SGD optimizer with the initial learning rate of 10$^{-4}$ and reduce it if the validation loss has not decreased for 20 consecutive epochs. The hyperparameter $\tau$ is set to 0.07. To leverage the auxiliary role of the contrastive learning loss, we test various $\lambda$ values and achieve the best results with $\lambda$ set to 0.1 on the IEMOCAP dataset. We search the best $k$ for the $k$NN model in the range $[1,\ldots, 32]$, as well as the best $\alpha \in (0,1)$ for each fold. We use weighted accuracy (WA) and unweighted accuracy (UA) as metrics to evaluate the performance of our proposed methods.

\subsection{Results and Discussions}
\begin{figure}[t]
	\centering
	\begin{subfigure}{0.325\linewidth}
		\centering
		\includegraphics[width=1\linewidth]{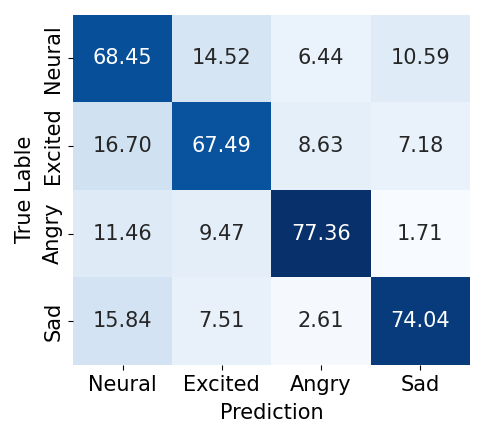}
		\caption{}
		\label{conf1}
	\end{subfigure}
	\centering
	\begin{subfigure}{0.325\linewidth}
		\centering
		\includegraphics[width=1\linewidth]{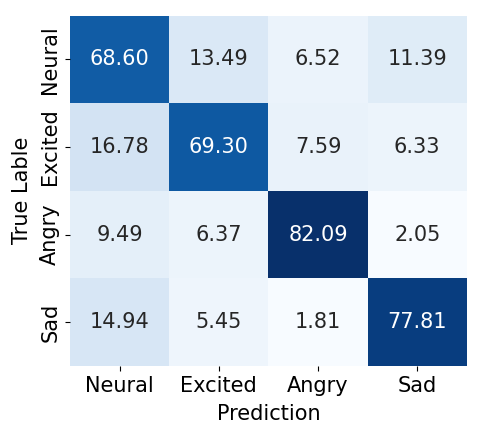}
		\caption{}
		\label{conf2}
	\end{subfigure}
	\centering
	\begin{subfigure}{0.325\linewidth}
		\centering
		\includegraphics[width=1\linewidth]{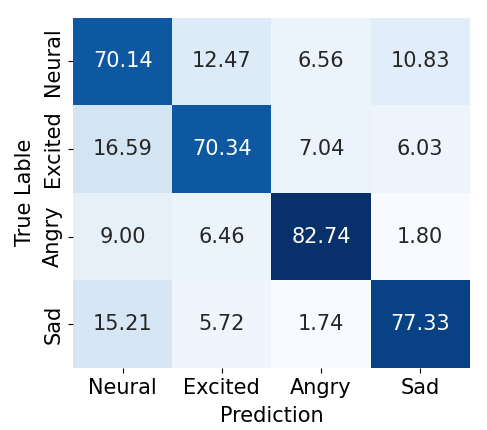}
		\caption{}
		\label{conf3}
	\end{subfigure}
   	\caption{Confusion matrices of the results during different stages. (a): Results after pre-training with wav2vec2.0. (b): Results after fine-tuning with our new loss function. (c): Results after interpolating model outputs with a $k$NN model in the inference stage. Numbers represent UA for each emotion.}
	\label{conf}
\end{figure}

Table \ref{tab:all results} shows the main results of our proposed methods at different stages, including pre-training, fine-tuning, and inference stages. Note that we utilize wav2vec2.0 as the pre-trained model for feature extraction during the pre-training stage. To evaluate the pre-training stage, the cross-entropy loss is used as the baseline (the S1 row in table \ref{tab:all results}) for fine-tuning the model for emotion prediction. We can observe a gradual improvement in performance throughout the SER pipeline as each method is applied, culminating in the best results achieved with the comprehensive approach. Particularly, the comprehensive approach outperforms the baseline system relatively by 3.90\% and 4.59\% on WA and UA, respectively. Figure \ref{conf} also demonstrates the same results with confusion matrices, which show more detailed performance for each emotion category. Further details of the experiments are discussed below.

\begin{table}[th]

  \caption{Evaluation results on IEMOCAP at different stages of the whole SER pipeline.}
  \label{tab:all results}
  \centering
  \begin{tabular}{llll}
    \toprule
    \textbf{Stages} & \textbf{Main Method} & \textbf{WA(\%)} & \textbf{UA(\%)}\\
    \midrule
    S1 (Pre-training) & Wav2vec2.0 & 71.35 & 71.84\\
    S2 (Fine-tuning) & S1 + SCL & 73.32 & 74.45\\
    S3 (Inference) & S2 + $k$NN & \textbf{74.13} & \textbf{75.14}\\
    \bottomrule
  \end{tabular}
  \vspace{-6pt}
\end{table}

\subsubsection{Effectiveness of SCL loss during fine-tuning}
\begin{figure}[t]
	\centering
	\begin{subfigure}{0.49\linewidth}
		\centering
		\includegraphics[width=1\linewidth]{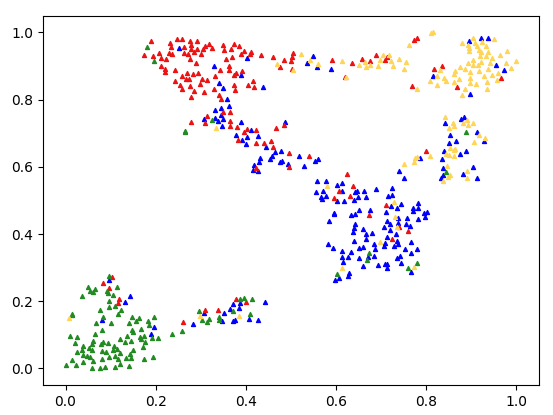}
		\caption{}
		\label{tsne2}
	\end{subfigure}
	\centering
	\begin{subfigure}{0.49\linewidth}
		\centering
		\includegraphics[width=1\linewidth]{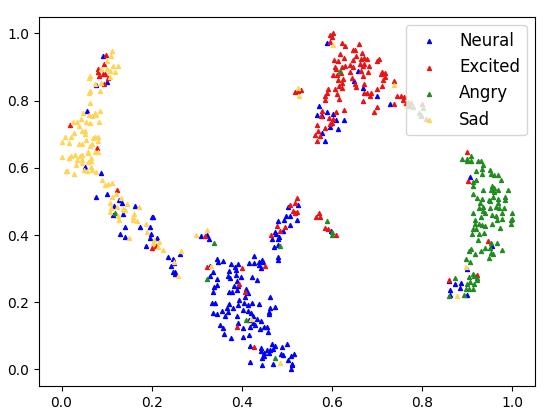}
		\caption{}
		\label{tsne3}
	\end{subfigure}
   	\caption{Visualization of the distributions about feature representations when fine-tuning with different strategies. (a): Outputs after fine-tuning with only cross-entropy loss. (b): Outputs after fine-tuning with the help of SCL.}
	\label{TSNE}
 \vspace{-6pt}
\end{figure}

To demonstrate the effectiveness of the new loss function, we visualize the outputs of the intermediate layer using the t-SNE technique \cite{2008Visualizing}. Figure \ref{TSNE}(a) and figure \ref{TSNE}(b) visualize the distributions of feature representations obtained by fine-tuning with only cross-entropy loss and our new loss function respectively. It could be found that after fine-tuning with a normal strategy, the model gets a basic ability to distinguish emotions. However, there still exist overlaps of certain emotions. After fine-tuning with the help of SCL loss, the boundaries between different emotions become clearer, especially for happy and neutral emotions, which are recognized as more difficult emotions to distinguish in previous researches \cite{chen20183}. The above findings suggest that our new learning objective combining SCL loss improves the quality of emotional representation effectively.

\subsubsection{Performances with different data augmentations}
In this paper, we also explore the model performances with different data augmentations in the fine-tuning stage, including adding noise, changing volume, adding reverberation, changing pitch, and mixing the above methods. The experimental results of various augmentations are shown in table \ref{tab:dataaug}.

It could be found that mixing several methods is more effective for improving the classification ability of the model. A complex method leads to more differences between \textit{positives}, which could make the model learn more information and have a stronger ability to distinguish emotions in a complicated situation. Mixed data augmentation helps contrastive learning play a better role in the fine-tuning stage.

\begin{table}[th]
  \caption{Performances with different methods of data augmentations in the fine-tuning stage.}
  \label{tab:dataaug}
  \centering
  \begin{tabular}{lll}
    \toprule
    \textbf{Augmentation Methods} & \textbf{WA(\%)} & \textbf{UA(\%)}\\
    \midrule
    Noise & 72.02 & 73.23\\
    Volume & 72.72 & 73.97\\
    Reverberation & 72.64 & 73.81\\
    Pitch & 72.75 & 73.88\\
    \textbf{Mixed augmentations} & \textbf{73.32} & \textbf{74.45}\\
    \bottomrule
  \end{tabular}
  \vspace{-6pt}
\end{table}

\subsubsection{Effectiveness of interpolation in the inference stage}
In the inference stage, we employ a $k$NN model to interpolate the fine-tuned model. To investigate whether SCL improves the sample distances and thus benefits $k$NN, we apply the $k$NN model to two fine-tuned models with different strategies: one fine-tuned with cross-entropy loss only (Loss w/o SCL), and the other with the incorporation of SCL (Loss w/ SCL). The results are reported in table \ref{tab:knn}.

\begin{table}[th]
  \caption{Performances before and after introducing $k$NN algorithm in the inference stage with different fine-tuning strategies}
  \label{tab:knn}
  \centering
  \begin{tabular}{lllll}
    \toprule
    \textbf{Fine-tuning} & \textbf{Inference} & \textbf{WA(\%)} & \textbf{UA(\%)} &\\
    \midrule
    Loss w/o SCL & w/o $k$NN & 71.35 & 71.84\\
    Loss w/o SCL & w/ $k$NN & 71.85 & 72.61 & +\\
    \midrule
    Loss w/ SCL & w/o $k$NN &  73.32 & 74.45\\
    Loss w/ SCL & w/ $k$NN & \textbf{74.13} & \textbf{75.14} & +\\
    \bottomrule
  \end{tabular}
  \vspace{-6pt}
\end{table}
Our experiments reveal that applying a $k$NN model to interpolate the original outputs leads to improved performance, regardless of the fine-tuning strategy used. By leveraging the feature information obtained during training, the $k$NN model is able to correct predictions made by the original model. 

Furthermore, the improvements in WA provide an intuitive reflection of the total number of corrected samples, highlighting the promising correction capability of the $k$NN model in our work. As shown in table \ref{tab:knn}, after fine-tuning with SCL, the improvement of WA is more obvious, suggesting that $k$NN could play a better role in correcting the predictions under this condition. 
Additionally, in the embedding space generated by SCL, the predictions of some samples have already been corrected, indicating the presence of hard samples with incorrect predictions. In such cases, $k$NN can still leverages the improved sample distances to enhance the model's classification ability without any additional training.

\subsection{Comparison to SOTA Approaches}
Table \ref{tab:sota} presents a comparison of our proposed method with recent state-of-the-art (SOTA) approaches. The results show that our proposed method, which applies improvements throughout the whole SER pipeline, achieves a relative improvement of 2.28\% and 1.13\% on WA and UA, respectively, compared to the best results of the SOTA approaches. These findings demonstrate the effectiveness of our approach.

\begin{table}[th]
  \caption{Performance comparison of our proposed methods with SOTA approaches on IEMOCAP.}
  \label{tab:sota}
  \centering
  \begin{tabular}{lll}
    \toprule
     \textbf{Model} & \textbf{WA(\%)} & \textbf{UA(\%)}\\
    \midrule
    Zou et al. \cite{9747095} & 71.64 & 72.70 \\ 
    Lu et al. \cite{lu2020speech} & 71.72 & 72.56\\ 
    Hu et al. \cite{hu22e_interspeech} & 69.31 & 70.11\\  
    Hu et al. \cite{hu22c_interspeech} & 72.48 & 67.72\\  
    Wav2vec2.0-PT \cite{pepino21_interspeech} & -- & 66.30\\ 
    Wav2vec2.0 P-TAPT \cite{chen2021exploring} & -- & 74.30\\ 
    \midrule
    \textbf{Ours} & \textbf{74.13} & \textbf{75.14}\\
    \bottomrule
  \end{tabular}
  \vspace{-6pt}
\end{table}

\section{Conclusions}

In this paper, we proposed a comprehensive framework to improve the performance of SER throughout its lifecycle, including pre-training, fine-tuning, and inference stages. We conducted a series of experiments which proved the effectiveness and necessity of our proposed methods in each stage. In comparison to state-of-the-art results, our proposed methods showed significant improvements on both WA and UA.

\section{Acknowledgments}
This work was supported by the National Key R$\&$D Program of China (Grant No.2022ZD0116307) and NSF China (Grant No.62271270).

\bibliographystyle{IEEEtran}

\bibliography{mybib}

\end{document}